# Identification of Parton Pairs in a Dijet Event and Investigation of Its Effects on Dijet Resonance Search

Sertac Ozturk[1,2]

[1] Gaziosmanpasa University, Department of Physics, 60150, Tokat, Turkey
[2] The University of Iowa, Department of Physics and Astronomy, Iowa City, IA 52242, USA

e-mail: sertac.ozturk@gop.edu.tr, sertac.ozturk@cern.ch

Being able to distinguish parton pair type in a dijet event could significantly improve the search for new particles that are predicted by the theories beyond the Standard Model at the Large Hadron Collider. To explore whether parton pair types manifesting themselves as a dijet event could be distinguished on an event-by-event basis, I performed a simulation based study considering observable jet variables. I found that using a multivariate approach can filter out about 80% of the other parton pairs while keeping more than half of the quark-quark or gluon-gluon parton pairs in an inclusive QCD dijet distribution. The effects of event-by-event parton pair tagging for dijet resonance searches were also investigated and I found that improvement on signal significance after applying parton pair tagging can reach up to 4 times for gluon-gluon resonances.

## 1. Introduction

A huge amount of jets are produced in the Large Hadron Collider (LHC) at CERN; mostly from QCD parton-parton scattering. The outgoing partons are observed as particle jets in a detector. The type of parton the jet came from can be determined using some properties of the initiating parton. This information is useful for many physics analyses.

Identification of a b-quark [1] is necessary for reconstruction of many physics processes, such as top quark decay or Higgs boson decay. Compared with the light quarks, the b-quark is heavier and has a long lifetime. In addition, it's decays are mostly semileptonic. These properties cause some differences in observable variables, such as the impact parameter of charged tracks, reconstruction of the secondary vertex, and the absence or presence of a lepton inside the jet.

Distinguishing quark-initiated jets from gluon-initiated jets is possible due to several differences between quarks and gluons, which lead to differences in observable variables in jets [2, 3]. Quark/gluon jet separation is very useful to search for particular physics processes. For example, many supersymmetric models produce quark jets with missing transverse energy. In addition, vector boson fusion produces forward jets, which are always quark-initiated.

There are many theories to extend the Standard Model that predict short-lived massive new particles coupling to quarks and gluons. These new particles are produced as narrow resonances, decaying to dijets and may show up as a bump in the invariant dijet mass distribution, emerging from the steeply falling distribution of the QCD dijet background. The decay of a dijet resonance can leave a final state consisting of quark-quark, quark-gluon and gluon-gluon parton pairs, which manifest themselves as a dijet event. A review of experimental searches for dijet resonances can be seen in reference [4].

Dijet resonance searches are one of the high priority physics topics of ATLAS [5-8] and CMS [9-12] experiments. In CMS experiment, dijet resonance signals are classified according to parton pair types and 95% cross section upper limits are set separately on each parton pair type. While the same observed dijet invariant mass spectrum is used, different signal shapes are chosen based on simulations of dijet resonances.

The search for b-tagged dijet resonance [11], exploiting a b-tagging algorithm from reconstructed objects to select b-quark originated jets, is performed categorizing the invariant dijet mass distribution with zero, one and two b-jets tagged. This additional requirement results in an increased search sensitivity to narrow resonances preferentially decaying into pairs of b quarks.

Similarly, several advantages may be gained by categorizing the dijet mass distribution based on parton pair tagging. It allows classifying dijet events to be able to search for particular parton final states, which may increase the signal sensitivity significantly. In the case of existence of a signal in a dijet mass spectrum, parton pairs tagging of the signal help us to understand the theoretical origin of the signal.

In this article, mainly two researches are performed. First, the performance of parton pair separation in a dijet event is investigated in Section 2. Second, the effects of parton pair separation for dijet resonance searches at LHC are studied in Section 3.

## 2. Identification of Parton Pairs

A dijet event may consist of three types of parton pair final states, which are quark-quark, quark-gluon and gluon-gluon. Two different theoretical models of dijet resonance are used to obtain the three types of final state parton pairs. The Randall-Sundrum Graviton model [13] with $k/M_{Pl}$=0.1 is used to obtain quark-quark and gluon-gluon parton pairs, where $M_{Pl}$ is the reduced Planck scale and $k$ is the unknown curvature scale of the extra dimension. The excited quark model [14, 15], with the compositeness scale set equal to the excited quark mass, is used to obtain quark-gluon parton pairs. Only light quarks in resonance decays are considered, since it was shown in [16, 17] that b-quark jets have properties that are more similar to gluon jets than to light quark jets. QCD could also be used to generate dijet events to obtain different types of parton pairs. I see that there is no effect on the results of parton pair separation whether dijet events are produced in s-channel (via resonances) or mainly t-channel (via QCD).

The events are generated using PYTHIA v8.180 [18] considering 14 TeV proton-proton collisions at various resonance masses, from 1 TeV to 7 TeV, to have a wide spectrum of the jets $p_T$. The jets are reconstructed using FASTJET v3.5.5 [19]. The charged particles with $p_T$<1 GeV are rejected and not used in the jet reconstruction to emulate detector effects. Anti-*kt* jet reconstruction algorithm [20] with cone size *R*=0.5 is used and two leading jets in an event are considered. Both leading jets were required to satisfy the $\eta$ cuts which are $|\eta_1, \eta_2| < 2.5$ and $|\Delta\eta| < 1.3$, which are the standard kinematic cuts of ATLAS and CMS for dijet resonance searches. This kinematic selection is optimal for enhancing isotropic signals over the QCD dijet background.

To be able to distinguish parton pair types in a dijet event, we can use several observable jet variables, which show variation due to various differences between quarks and gluons. Basically, there are two types of observable variables for quark/gluon jet tagging: discrete variables and continuous variables. Charged particle multiplicity is an example of a discrete variable which is sensitive to the color factor of quarks ($C_F$=4/3) and gluons ($C_A$=3). In addition, number of subjets above a particular $p_T$ threshold, $p_T$ fraction of $N^{th}$ subjet, average distance from jet axis, and charge-weighted $p_T$ sum of tracks are the other examples of discrete variables. The continuous observables include jet mass, jet broadening and jet angularities. The details of these observables are reported in reference [3].

Usually, using several variables at the same time can improve tagging results significantly. TMVA v4.1.3 package [21] part of ROOT v34.02 [22] is used to combine the observable variables. The combination of the following variables from two leading jets is found to be optimal for distinguishing parton pairs: Jet Mass/$p_T$ with *R*=0.3, $p_T$ fraction of jet with *R*=0.2, charged particle multiplicity and jet girth [3, 23], which is defined as

$$g = \frac{1}{p_T^{jet}} \sum_{i \in jet} p_T^i \sqrt{r_i}$$

where $r_i = \sqrt{(\Delta\eta_i)^2 + (\Delta\phi_i)^2}$. Since charged particle multiplicity and jet mass are proportional to the color factor, charged particle multiplicity and jet Mass/$p_T$ ratio of gluon jets are greater than quark jets. Quark jets are narrower than gluon jets and jet energy fraction near the center of gluon jets is less than quark jets.

Figure 1 shows the considered jet observable distributions of two leading jets at the resonance mass of 3 TeV, for which jet $p_T$ varies between 100 GeV to 2 TeV. Similar distributions are also observed at the different values of resonance masses. The distributions show good separation between quark-quark and gluon-gluon final states. As expected, the distributions of quark-gluon fall between quark-quark and gluon-gluon.

Considering two leading jets, a total of eight jet variables is found as the inputs for an event. Boosted Decision Trees (BDT) method is chosen as a multivariate discriminator. Likelihood and Artificial Neural Networks methods are also considered. But it is found that BDT has the highest background rejection at any signal efficiency and it behaves more stable for all resonance masses than the other techniques. Figure 2 presents BDT output distributions for each type of parton pair separation. Since there are three types of parton pairs that can produce a dijet event, the contributions to background should come from the other two different parton pair types. For example, if we are interested in quark-quark parton pair separation, the background should be combination of quark-gluon and gluon-gluon parton pairs. The author notes that although these BDTs were trained and measured using parton pairs from dijet resonance Monte Carlo, performance on parton pairs coming from QCD events is expected to be similar.

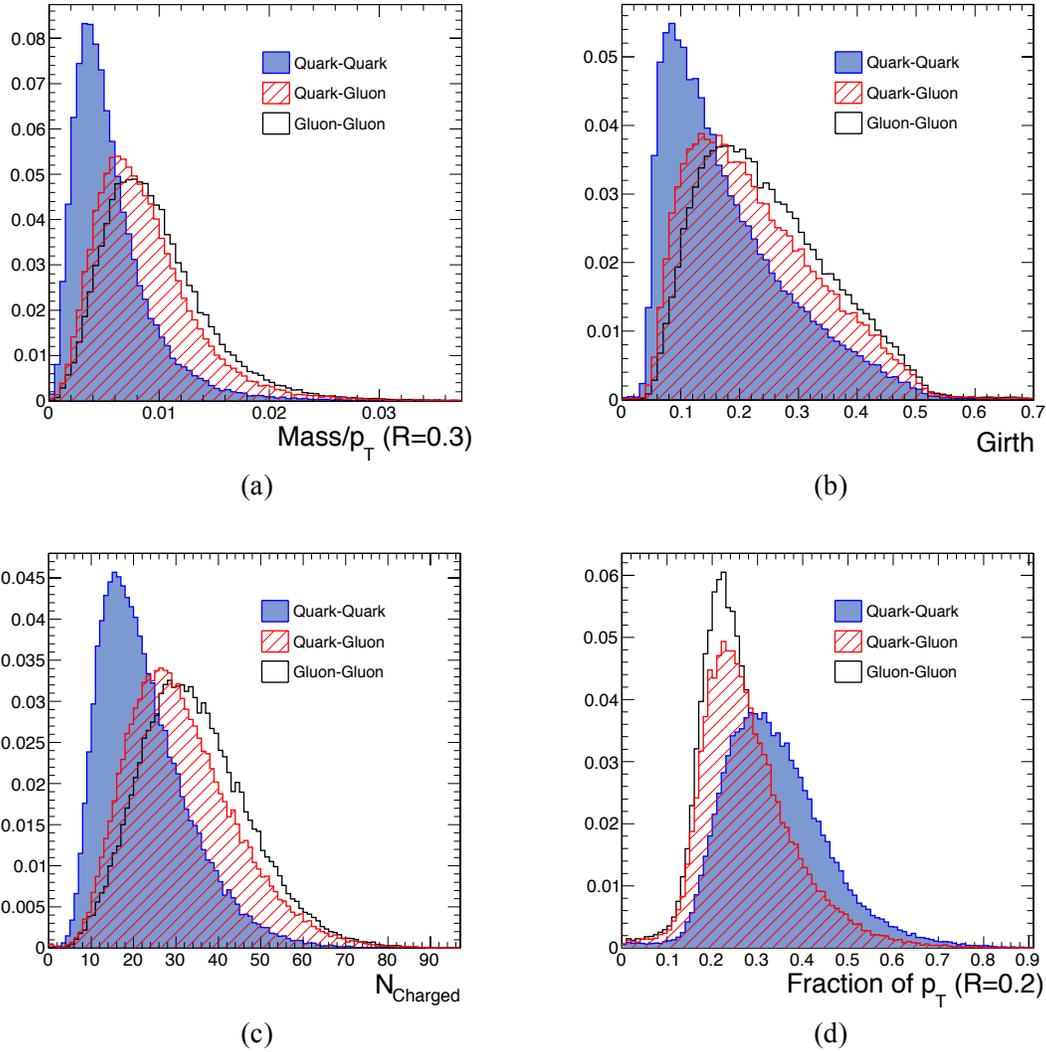

Figure 1. The distributions of selected variables from two leading jets: The ratio between jet mass and jet $p_T$ with $R$=0.3 in (a), Girth in (b), charged particle multiplicities in (c), and fraction of jet $p_T$ with R=0.2 in (d). Solid blue histograms present quark-quark final state, dashed red histograms show quark-gluon final state and empty black lined histograms present gluon-gluon final state at the resonance mass of 3 TeV.

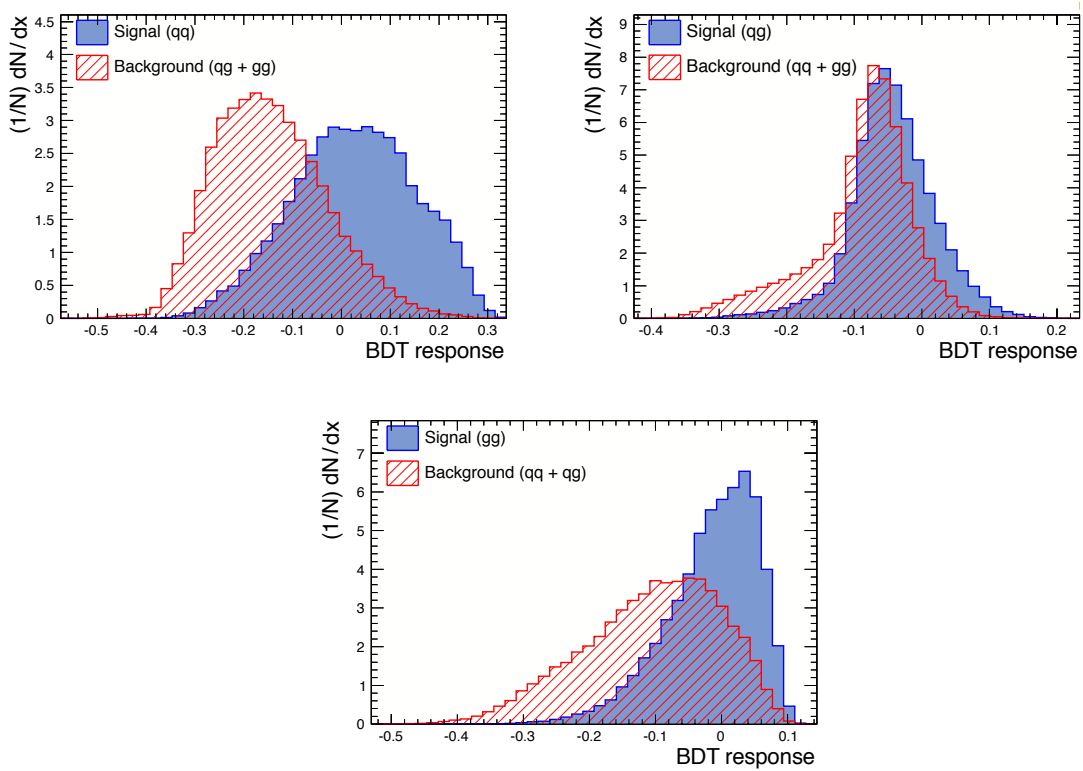

Figure 2. The output distribution of Boosted Decision Tree technique for each type of parton pair separation. Blue solid histograms show signal of a particular parton pair and red hatched ones present background that comes from other two types of parton pairs.

The ROC (Receiver operating characteristic) curves that describe the signal efficiency versus background rejection are shown in Figure 3 for each type of pairs. The best background rejection in a given signal efficiency point are obtained for quark-quark parton pair. Since the jet variables of quark-gluon parton pair fall between quark-quark and gluon-gluon decays, it is the hardest parton pair type to distinguish from other parton pair types. As an example, an inclusive QCD dijet distribution can be considered. Figure 3 shows that 90% of events, which are produced in quark-gluon and gluon-gluon parton pairs, can be rejected in the inclusive QCD dijet mass distribution while keeping around 60% of quark-quark parton pair events.

The ROC curve in Figure 3 is obtained using the jet variables at the resonance mass of 3 TeV. For the all resonance masses corresponding to different dijet mass window, the best separation is obtained for quark-quark parton pair, while the worst separation is obtained for quark-gluon. Variation in the ROC curves is small when using different resonance masses. The variation of background rejection at a given signal efficiency is up to 7% for quark-quark decays, while it is only 2% for gluon-gluon decays, for resonance masses between 1 and 7 TeV.

The background rejection at a given signal efficiency increases when resonance mass increases since the jet energy also increases. Because performance of quark/gluon jet separation is better at high jet $p_T$, parton pair separation can be performed better at higher masses.

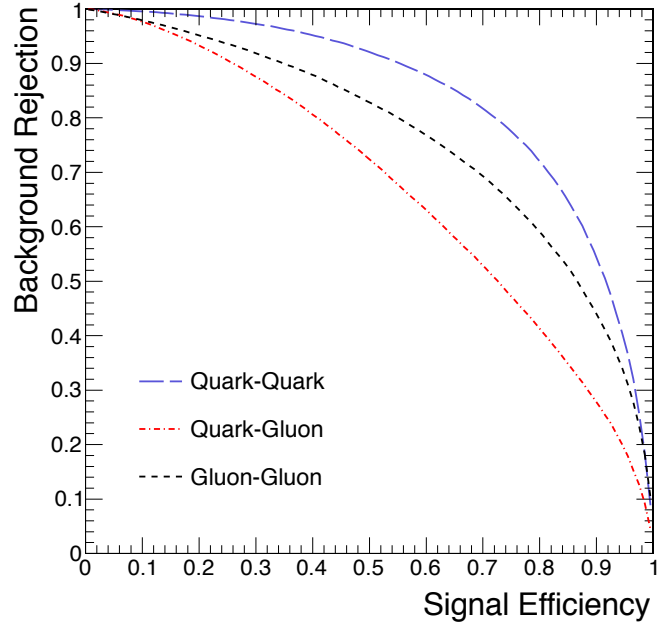

Figure 3. ROC curves of each type o parton pairs. The greatest background rejection is achieved for quark-quark parton pair tagging.

**3. Improvement in Dijet Resonance Search**

In the previous chapter, it is presented that parton pair tagging can be performed using jet observable variables. It allows to classify a dijet mass spectrum according to parton pair types. The sensitivity to dijet resonance decaying to a specific parton pair should increase after applying a parton pair tagging method, because parton pair tagging suppresses the QCD background. In order to study improvement in dijet resonance search, QCD dijet events are generated using PYTHIA v8.180 to be used as background.

Various discriminator points can be selected to operate parton pair tagging in an event using BDT output distributions shown in Figure 2. The selection of discriminator cut points can be done considering signal efficiency or background rejection. I select several discriminator points based on signal efficiency in a ROC curve. The ROC curves and BDT outputs at the resonance mass of 3 TeV is used because it gives relatively wide variation in jet $p_T$.

The use of parton pair tagging can affect the signal modeling in two ways. First, it can potentially change the shape of the signal. Second, it reduces the signal efficiency. I find that the application of parton pair tagging does not appreciably change the resonance shapes as long as the signal efficiency is sufficiently high. But at high resonance mass, using tighter cuts which correspond to below 40% of signal efficiency, resonance masses are effected due to low mass tail coming from final state radiation and parton distribution function effect. Another impact of parton pair tagging on signal modeling is its affect on the signal efficiency. The signal efficiencies vary ±10% over all the considered resonance masses.

Similarly, parton pair tagging is applied to QCD background to classify dijet mass distribution in different parton pair category. After parton pair tagging is applied to the inclusive QCD background, reasonable falling QCD background spectrums are obtained for each type of parton pairs.

Figure 4 (a) shows improvement of the signal significance as a function of resonance mass at 80% signal efficiency cut selection. The searching mass windows are selected between ±σ variations of mean value for each resonance masses. The improvement is not notable for quark-quark and quark-gluon dijet resonances, but it may help to determine the origin of a dijet resonance signal. As the value of resonance masses increase, improvement of quark-quark signals decrease, while it is almost constant for quark-gluon. The greatest effect of parton pair tagging to improve and identify resonance signals is obtained for gluon-gluon resonances. This is because high mass QCD distribution is dominated by quark-quark final state, which is most easily distinguished from the gluon-gluon final state. As the value of resonance masses increase, improvement of gluon-gluon signals increase.

Figure 4 (b) presents the effects of the discriminator selection on signal significances at the resonance mass of 5 TeV. Choice of the discriminator points almost does not change the improvement of the signal for quark-quark and quark-gluon resonances. But the gluon-gluon signal sensitivity increases when using tighter discriminator cut.

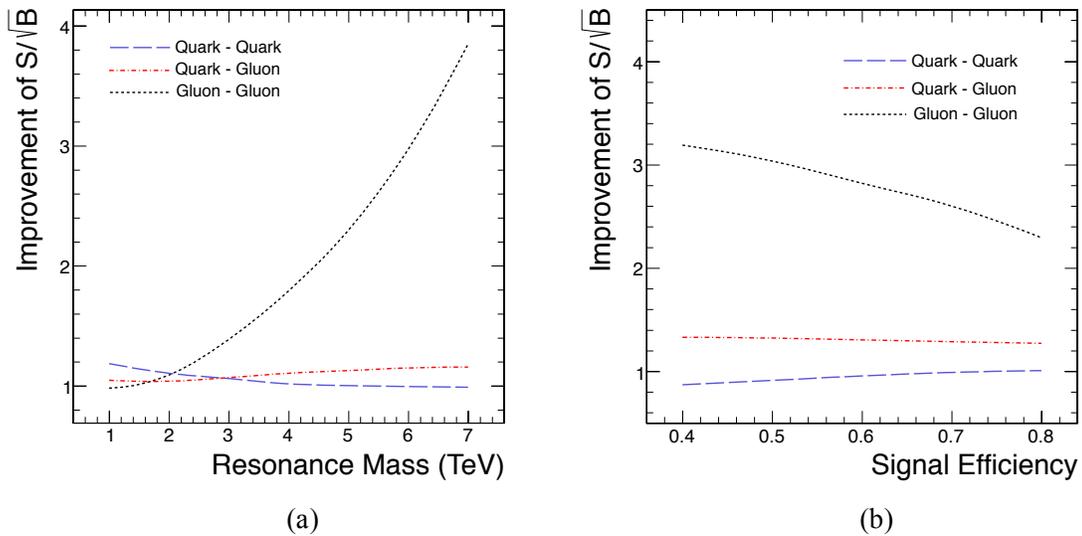

(a)          (b)

Figure 4. (a) Improvement of signal significance ratio as a function of resonance mass after applying parton tagging. The tagging discriminator is chosen at 80% signal efficiency. (b) Improvement of signal significance as a function of signal efficiency cut at the resonance mass of 5 TeV. In both plots, black dashed line shows gluon-gluon resonances, red dashed line present quark-gluon resonances and purple dashed line shows quark-quark resonances.

## 4. Conclusion

LHC is going to start taking data at higher collision energy in 2015 and it is going to allow us to explore the higher energy regime. Undoubtedly, dijet resonance search is going to be one of the top research topics of CMS and ATLAS. This paper is addressed two main studies. First, it is presented that parton pair tagging may be achieved in a dijet event, especially for quark-quark and gluon-gluon parton pairs. Second, a new approach is studied using parton pair tagging to improve the signal sensitivity of searches for dijet resonances. It is seen that use of parton pair tagging has the potential to enhance sensitivity to many models of dijet resonances, and I find it significantly increase the signal significance for gluon-gluon resonances. Parton pair tagging may also help identify the parton pair source of the resonance signals, which would help to understand the theoretical model behind any resonance discovered.

**Conflict of Interests**

The author declares that there is no conflict of interests regarding the publication of this paper.

**References**

[1] The CMS Collaboration, "Identification of b-quark jets with the CMS experiment", *Journal of Instrumentation*, vol. 8, Article ID P04013, 2013.

[2] J. Gallicchio and M. D. Schwartz, "Quark and Gluon Tagging at the LHC", *Physical Review Letters*, vol. 107, Article ID 172001, 2011.

[3] J. Gallicchio and M. D. Schwartz, "Quark and Gluon Jet Substructure", arXiv:1211.7038v2, 2011

[4] R. M. Harris and K. Kousouris, "Searches for Dijet Resonances at Hadron Colliders", *International Journal of Modern Physics A*, vol. 26, no. 30-31, pp. 5005–5055, 2011.

[5] G. Aad, B. Abbott, J. Abdallah et al., "Search for new particles in two-jet final states in 7 TeV proton-proton collisions with the ATLAS detector at the LHC", *Physical Review Letters*, vol. 105, no. 16, Article ID 161801, 19 pages, 2010.

[6] G. Aad, B. Abbott, J. Abdallah et al., "A search for new physics in dijet mass and angular distributions in *pp* collisions at √s = 7 TeV measured with the ATLAS detector", *New Journal of Physics*, vol. 13, no. 5, Article ID 053044, 2011

[7] G. Aad, B. Abbott, J. Abdallah et al., "Search for new physics in the dijet mass distribution using 1 fb$^{-1}$ of *pp* collision data at $\sqrt{s}$ = 7 TeV collected by the ATLAS detector", *Physics Letters B*, vol. 708, no. 1-2, pp. 37–54, 2012.

[8] G. Aad, T. Abajyan, B. Abbott et al., "ATLAS search for new phenomena in dijet mass and angular distributions using pp collisions at $\sqrt{s}$ = 7 TeV", *Journal of High Energy Physics*, vol. 2013, p. 029, 2013.

[9] S.Chatrchyan, V.Khachatryan, A.M.Sirunyan et al.,"Search for resonances in the dijet mass spectrum from 7 TeV pp collisions at CMS", *Physics Letters B*, vol. 704, no. 3, pp. 123–142, 2011.

[10] S.Chatrchyan, V.Khachatryan, A.M.Sirunyan et al.,"Search for microscopic black holes in pp collisions at $\sqrt{s}$ = 7 TeV", *Journal of High Energy Physics*, vol. 2012, article 61, 2012.

[11] S. Chatrchyan, V. Khachatryan, A. M. Sirunyan et al., "Search for narrow resonances and quantum black holes in inclusive and b-tagged dijet mass spectra from pp collisions at $\sqrt{s}$ = 7 TeV", *Journal of High Energy Physics*, vol. 2013, article 13, 2013.

[12] S. Chatrchyan, V. Khachatryan, A. M. Sirunyan et al., "Search for narrow resonances using the dijet mass spectrum in *pp* collisions at $\sqrt{s}$ = 8 TeV", *Physical Review D*, vol. 87, no. 11, Article ID 114015, 16 pages, 2013.

[13] L. Randall and R. Sundrum, "An alternative to compactification", *Physical Review Letters*, vol. 83, no. 23, pp. 4690–4693, 1999.

[14] U. Baur, I. Hinchliffe, and D. Zeppenfeld, "Excited quark production at Hadron colliders", *International Journal of Modern Physics A*, vol. 2, p. 1285, 1987.

[15] U. Baur, M. Spira, and P. M. Zerwas, "Excited-quark and -lepton production at hadron colliders", *Physical Review D*, vol. 42, no. 3, pp. 815–824, 1990.

[16] D. Buskulic, D. Casper, I. de Bonis et al., "Quark and gluon jet properties in symmetric three-jet events", *Physics Letters B*, vol. 384, no. 1–4, pp. 353–364, 1996.

[17] O. Biebel, "A comparaison of b and uds quarks to gluon jets", in *Proceedings of the 9th Annual Meeting of the Division of Particles and Fields of the American Physical Society*, pp. 354–356, Minneapolis, MN, USA, 1996, SPIRES Conference C96/08/11.1.

[18] T. Sjöstrand, S. Mrenna, and P. Skands, "A brief introduction to PYTHIA 8.1", *Computer Physics Communications*, vol. 178, no. 11, pp. 852–867, 2008.

[19] M. Cacciari and G. P. Salam, "Dispelling the N$^3$ myth for the kt jet-finder", *Physics Letters B*, vol. 641, no. 1, pp. 57–61, 2006.

[20] M. Cacciari, G. P. Salam, and G. Soyez, "The Anti-k(t) jet clustering algorithm", *Journal of High Energy Physics*, vol. 2008, article 063, 2008.

[21] A. Hoecker, J. Stelzer, P. Speckmayer et al., TMVA Toolkit for Multivariate Data Analysis with ROOT, http://tmva.sourceforge.net/.

[22] R. Brun and F. Rademakers, "ROOT—an object oriented data analysis framework", *Nuclear Instruments and Methods in Physics Research A*, vol. 389, no. 1-2, pp. 81–86, 1997, http://root.cern.ch/.

[23] J. Gallicchio, J. Huth, M. Kagan, M. D. Schwartz, K. Black, and B. Tweedie, "Multivariate discrimination and the Higgs+W/Z search", *Journal of High Energy Physics*, vol. 2011, no. 4, article 069, 2011.